\documentclass[twocolumn]{emulateapj} 
\newcommand{\vek}[1]{\mbox{\boldmath$#1$}}

\shorttitle{SUCCESSIVE MERGING OF PLASMOIDS AND FRAGMENTATION}
\shortauthors{Karlick\'y and B\'arta}

\begin{document}

\title{SUCCESSIVE MERGING OF PLASMOIDS AND FRAGMENTATION IN FLARE
CURRENT SHEET AND THEIR X-RAY AND RADIO SIGNATURES}

\author{Marian Karlick\'y\altaffilmark{1}
  and Miroslav B\'arta\altaffilmark{1,2}}

\affil{$^1$Astronomical Institute of the Academy of Sciences
  of the Czech Republic, CZ-25165 Ond\v{r}ejov, Czech Republic\\
  $^2$Max Planck Institute for Solar System Research,
  D-37191 Katlenburg-Lindau, Germany}

\email{karlicky@asu.cas.cz}

\begin{abstract}
Based on our recent MHD simulations, first, a concept of the successive merging
of plasmoids and fragmentation in the current sheet in the standard flare model
is presented. Then, using a 2.5-D electromagnetic particle-in-cell model with
free boundary conditions, these processes were modelled on the kinetic level of
plasma description. We recognized the plasmoids which mutually interacted and
finally merged into one large plasmoid. Between interacting plasmoids further
plasmoids and current sheets on smaller and smaller spatial scales were formed
in agreement with the fragmentation found in MHD simulations. During
interactions (merging - coalescences) of the plasmoids the electrons were very
efficiently accelerated and heated. We found that after a series of such
merging processes the electrons in some regions reached the energies relevant
for the emission in the hard X-ray range. Considering these energetic electrons
and assuming the plasma density 10$^{9}$--10$^{10}$ cm$^{-3}$ and the source
volume as in the December 31, 2007 flare (Krucker at al. 2010), we computed the
X-ray spectra as produced by the bremsstrahlung emission process. Comparing
these spectra with observations, we think that these processes can explain the
observed above-the-loop-top hard X-ray sources. Furthermore, we show that the
process of a fragmentation between two merging plasmoids can generate the
narrowband dm-spikes. Formulas for schematic fractal reconnection structures
were derived. Finally, the results were discussed.
\end{abstract}

\keywords{Plasmas --- Solar flares --- Acceleration of particles}

\section{INTRODUCTION}

It is commonly accepted that plasmoids play a very important role in the
magnetic field reconnection in solar flares. Their importance, for the first
time, was recognized by \citet{OhyamaShibata1998}. 
In the 1992 October~5 flare,
observed in soft X-rays by {\it Yohkoh} satellite, they analyzed the plasmoid
which was ejected during the impulsive phase upwards into the corona. Studying
the same flare, \citet{Kliemetal2000} showed that this plasmoid ejection was
associated with the drifting pulsating structure (DPS) on
radiowaves. They proposed the model 
of this radio emission, which was further developed in the papers by 
\citet{Karlickyetal2002}, \citet{Karlicky2004}, \citet{KarlickyBarta2007},
\citet{Bartaal2008a}, and \citet{Karlickyetal2010}.
In this model, in the current sheet, due to tearing and coalescence
processes the plasmoids are formed. As shown by 
\citet{Drakeetal2005, Drakeetal2006},
\citet{Hoshino2005}, \citet{Pritchett2006, Pritchett2008}, and 
\citet{Karlicky2008} during these
processes electrons are very efficiently accelerated. The electrons are then
trapped in plasmoids, where they generate Langmuir waves, which through a wave
transformation produce the electromagnetic waves recorded on the radio spectrum
as DPSs. Due to limited range of plasma densities in the plasmoid the DPS is
generated in the limited range of frequencies. In the vertical current sheet
the plasmoids move upwards or downwards or even stay without any motion in
dependance on a form of the surrounding magnetic field 
\citep{Bartaal2008a, Bartaal2008b}. Due to a preference of 
divergent magnetic field lines in the upward
direction, most of the plasmoids move upwards and corresponding DPSs drift
towards lower frequencies. Nevertheless, in some cases the plasmoids move
downwards and even interact with the underlying flare arcade as observed by
\citet{KolomanskiKarlicky2007} and \citet{Milliganetal2010}.

Recently, \citet{Okaetal2010} studied the electron acceleration by multi-island
coalescence processes in PIC model with periodic boundary conditions. They
found that the most effective acceleration process is during the coalescence of
plasmoids ("anti-reconnection"), see also \citet{Pritchett2008} 
and \citet{KarlickyBarta2007}.

Furthermore, \citet{ShibataTanuma2001} proposed that the current sheet,
stretched by a rising magnetic rope, is fragmented to smaller and smaller
plasmoids by the tearing mode instability in subsequently narrower and narrower
current sheets (cascading reconnection). This suggestion has been recently
further theoretically developed by \citet{Loureiro+:2007}
and \citet{Uzdensky+:2010}
into the theory of chain plasmoid instability. An advantage of this
concept is that it explains how very narrow current sheets with high current
densities (requested for the anomalous resistivity generation and fast
reconnection) are generated. Moreover, many X-points in this model give
sufficient volume for an acceleration of particles.

Besides this fragmentation described by \citet{ShibataTanuma2001}, 
\citet{Bartaal2010b} found a new fragmentation in the region between two merging
plasmoids using MHD simulations. This fragmentation is caused by the tearing
mode instability in the current sheet generated between these interacting
plasmoids and this process is repeated in smaller and smaller spatial scales.
This fragmentation is driven by a merging process of the plasmoids.

Considering all the above-mentioned processes, in the present paper, we focus
our attention to two processes: (a) successive merging of plasmoids to large
plasmoid and (b) fragmentation process between merging plasmoids. We selected
these processes because we think that the successive merging of plasmoids can
explain the above-the-loop-top hard X-ray source (as a large stationary
plasmoid). On the other hand, the fragmentation can explain the narrowband
dm-spikes. Because both these phenomena are generated by accelerated electrons,
in following simulations we use the particle-in-cell (PIC) model instead of MHD
models \citep[e.g.][]{Bartaal2008a, Bartaal2008b}.

The above-the-loop-top hard X-ray sources belong to the most discussed topics
in recent years. The well-known example of such a hard X-ray source is that
observed in the $\sim$30--50 keV energy range by \citet{Masudaetal1994}. 
However,
such events are very rare \citep{Tomczak2001, Petrosian2002, KruckerLin2008}.
Another very interesting example was published just recently by 
\citet{Kruckeretal2010}. They presented the hard X-ray source 
(with the energy up to
$\sim$80 keV) which was located 6 Mm above thermal flare loops. They derived
the upper limit of the plasma density and source volume as n$_e \sim$
8$\times$10$^{9}$ cm$^{-3}$ and $V \sim$ 8$\times$ 10$^{26}$ cm$^{3}$,
respectively. Just a relatively low plasma density in such hard X-ray sources
attracts attention of scientists. \citet{Kruckeretal2010} concluded that these
hard X-ray sources have to be close to the acceleration region and the
distribution function of electrons emitting hard X-rays is strongly non-thermal
or the plasma in the source is very hot (up to T$_e$ $\sim$ 200 MK). Several
ideas explaining these X-ray sources have been proposed, e.g. the magnetic or
turbulent trapping, and dense (collisionally thick) coronal sources
\citep[see][]{Fletcher1995,Jakimiec1998,VeronigBrown2004,ParkFleishman2010}

The narrowband dm-spikes (further spikes) belong to the most interesting radio
bursts due to exceptionally high brightness temperatures
(T$_{b}$~$\approx$~10$^{15}$~K) and short durations 
\citep[$\leq$~0.1~s, see the review by][]{Benz1986}. 
Their observational characteristics were described in
many papers 
\citep[e.g.][]{Slottje1981,Karlicky1984,Fuetal1985,StahliMagun1986,
Benzetal1982,ZlobecKarlicky1998,Meszarosova2003}.
On the other hand, the theoretical models
can be divided into two groups: a)~based on the plasma emission and
acceleration processes 
\citep{Kuijpersetal1981,Tajimaetal1990,Wentzel1991,BartaKarlicky2001},
and b)~based on the electron-cyclotron maser
\citep{Holmanetal1980,MelroseDulk1982,VlahosSharma1984,Wingleeetal1988,
Aschwanden1990,FleishmanYastrebov1994}.
To distinguish between
these two types of models polarization and harmonic structures of the spikes
have been also studied 
\citep{Gudel1990,GudelZlobec1991,KruckerBenz1994}.
Searching for a~characteristic bandwidth of individual spikes 
\citet{Karlicky1996,Karlickyetal2000}
found that the Fourier transform of the dynamic spectra of
spikes have a~power-law form with power-law indices close to~$-5/3$. Based on
these results \citet{BartaKarlicky2001} proposed that the spikes are
generated in turbulent reconnection outflows.

This paper is organized as follows: First, we present our model scenario and
explain the successive merging and fragmentation process. Then using a 2.5-D
PIC model we simulate these processes. The results are then used in the
interpretation of the above-the-loop-top hard X-ray sources and narrowband
dm-spikes.

\begin{figure}
\begin{center}
\includegraphics[width=8.5cm]{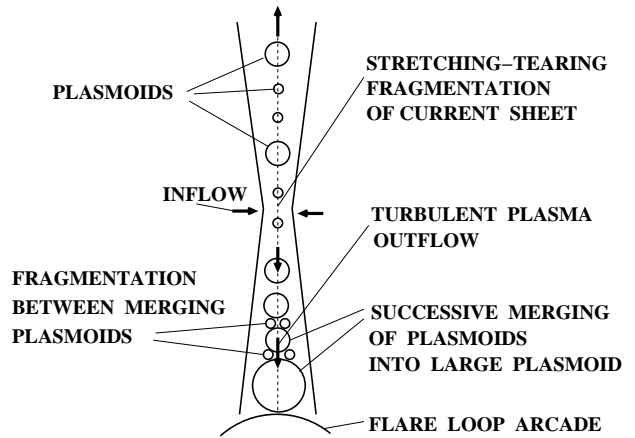}
\end{center}
    \caption{Model scenario.}
\label{figure1}
\end{figure}

\section{MODEL SCENARIO AND SIMULATION MODEL}

Fig.~\ref{figure1} shows our model scenario, which is based on the
'standard' CSHKP flare 
model \citep[e.g.][and references therein]{Magaraetal1996}. 
In the central part of
the current sheet, in agreement with \citet{ShibataTanuma2001}, we assume a
fragmentation of the current sheet (stretching-tearing fragmentation).
Furthermore, based on the new results of \citet{Bartaal2010b}, we propose
that the reconnection plasma outflow (which is downward oriented) accumulates
plasmoids in the region just above the flare arcade, where thus plasmoids can
interact efficiently. We think that in some cases a large plasmoid can be
formed here as a result of successive merging of plasmoids. On the other hand,
between the merging plasmoids new current sheets are formed and in these
current sheets one again further (but on smaller and smaller spatial scales)
plasmoids are generated (fragmentation between merging plasmoids). While the
first process of the successive merging we propose for the interpretation of
the above-the-loop-top hard X-ray sources, the second process is a very
promising process explaining the narrowband dm-spikes (see the following
sections \ref{sect31} and \ref{sect32}).

For simulation we used a 2.5-D (2D3V -- 2 spatial and 3 velocity components)
fully relativistic electromagnetic particle-in-cell model 
\citep{SaitoSakai2004,Karlicky2004}.
The system size is $L_x \times L_y$ = 600$\Delta$ $\times$
4000$\Delta$, where $\Delta$ (=1) is a grid size. The current sheet is
initiated along the line $x$ = 0$\Delta$, and its half-width is $L$ =
10$\Delta$. In this first study we consider the neutral current sheet, i.e. the
guiding magnetic field B$_z$ is zero. Thus, the initial magnetic field is:

\begin{eqnarray}
{\bf B} \equiv ({B_x, B_y, B_z}),\nonumber \\
B_y = -B_0~ {\rm for}~x < - L ,\nonumber \\ B_y = x B_0/L~{\rm for}~ -L
\leq~{\rm} x \leq~+L, \nonumber \\ B_y = B_0 ~ {\rm for}~x
> + L ,\nonumber \\ B_x = 0, B_z = 0. \nonumber
\end{eqnarray}
 
\begin{figure*}
\begin{center}
\includegraphics[width=17cm]{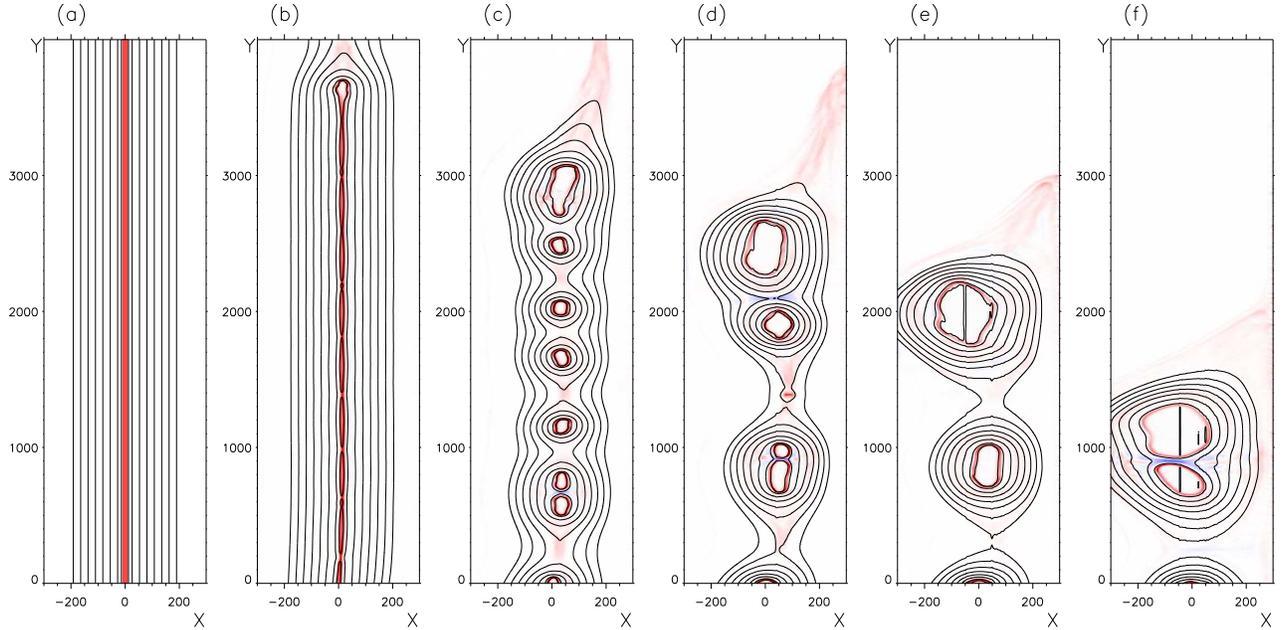}
\end{center}
\caption{The global view on magnetic field lines and corresponding
  current densities 
(reddish areas) in the $x-y$ computational plane at six different times: at the
initial state (a),  at $\omega_{pe} t$ = 1200
 (b), at $\omega_{pe} t$ = 2500 (c), at $\omega_{pe} t$ = 3500 (d),
 at $\omega_{pe} t$ = 4500 (e), and
 at $\omega_{pe} t$ = 6500 (f).
 The $x$ and $y$ coordinates are expressed in $\Delta$.
 The proton inertial length is 40 $\Delta$.}
\label{figure2}
\end{figure*}

The electron-proton plasma with the proton-electron mass ratio $m_p/m_e$=16
 is unrealistic, but taken here to
shorten computations. Nevertheless, the electron mass is low enough to separate
the dynamics of electrons and protons well. In each numerical cell located
outside of the current sheet, we initiated n$_0$ = 60 electrons and n$_0$ = 60
protons. In the current sheet the initial number density was enhanced just to
keep the pressure equilibrium. The initial electron temperature was taken to be
the same in the whole numerical box as T = 10 MK and the temperature of protons
was chosen the same as electrons. The plasma frequency is $\omega_{pe} \Delta
t$ = 0.05 ($\Delta t$ is the time step which equals to 1), the electron Debye
length is $\lambda_D = 0.6\ \Delta$, and the electron and proton inertial
lengths are $d_{e} = 10\ \Delta$ and $d_{i} = 40\ \Delta$, respectively. To
study successive coalescence processes among several plasmoids, we initiated a
formation of 10 plasmoids along the current sheet by a cosine perturbation of
the electric current density in the sheet; with the k-vector 
$k = 2\pi\cdot10/4000 = 0.0157\ \Delta^{-1}$ and the amplitude 
corresponding to the current density
{\bf j} given by the magnetic field in the current sheet 
($\vek{j=\nabla\times B}$).
We made computations with several initial values of the plasma $\beta$
parameter. Here, we present the results with $\beta$ = 0.07. It gives the ratio
of the plasma densities in the centrum and out of the current sheet as 15.3.
The free boundary conditions were used. All computations were performed on the
parallel computer OCAS (Ond\v{r}ejov Cluster for Astrophysical Simulations),
see http://wave.asu.cas.cz/ocas.

\section{RESULTS}

\begin{figure*}
\begin{center}
\includegraphics[width=17cm]{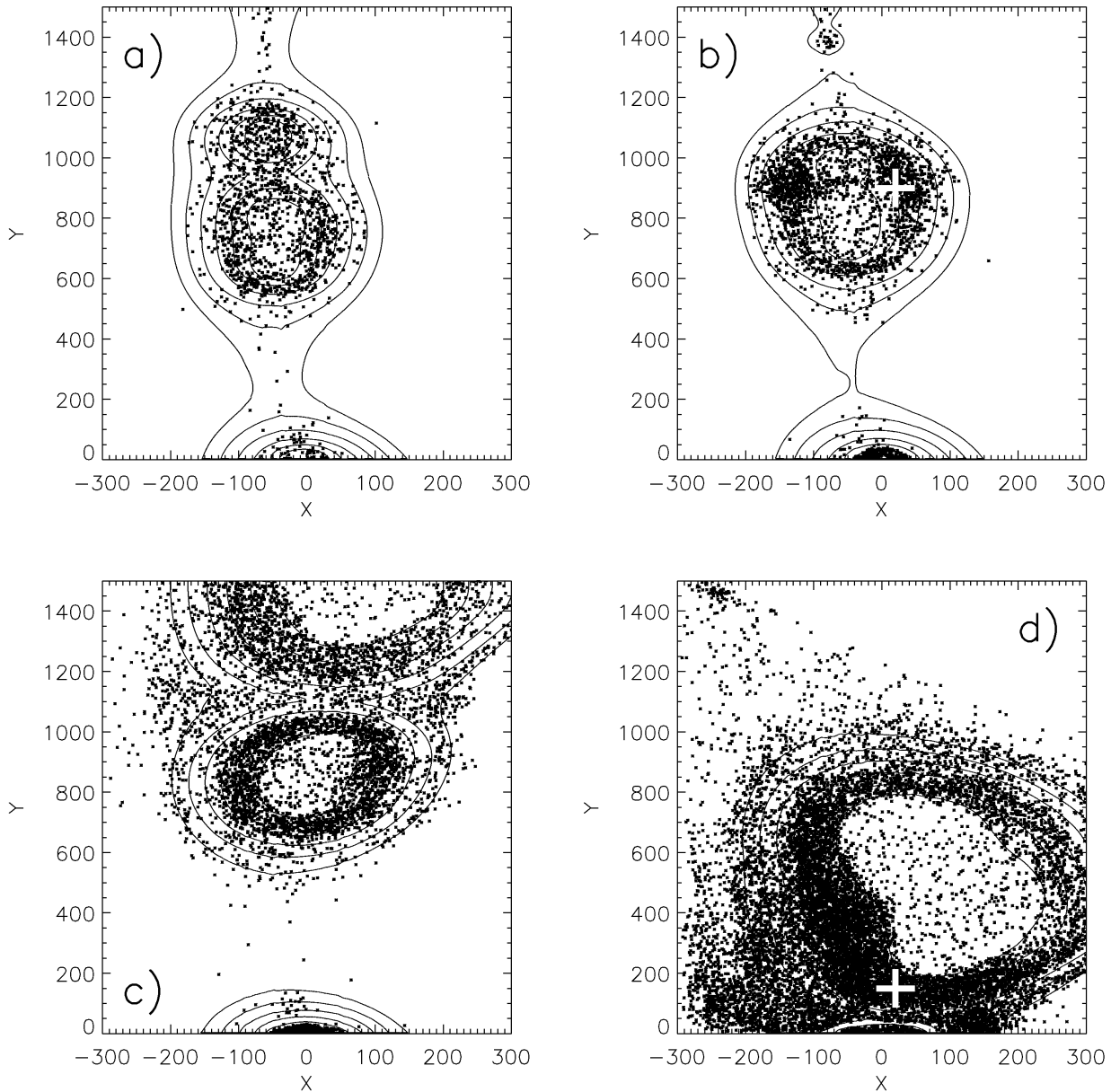}
\end{center}
    \caption{The detailed view on magnetic field lines in the $x-y$
    computational plane and 
    distribution of numerical electrons (points) having the energy
    greater than 40 keV 
    at four different times:
    at $\omega_{pe} t$ = 3200
 (a), at $\omega_{pe} t$ = 3500 (b), at $\omega_{pe} t$ = 5600 (c), and
 at $\omega_{pe} t$ = 7800 (d). The white crosses, located at b) x =
    20, y = 900, and at 
 d) x = 20, y = 150, show the regions where 
 the distribution functions and the X-ray spectra were computed 
 (see Figs.~\ref{figure4} and \ref{figure5}).
 The $x$ and $y$ coordinates are expressed in $\Delta$.}
\label{figure3}
\end{figure*}

\begin{figure}
\begin{center}
\includegraphics[width=8.5cm]{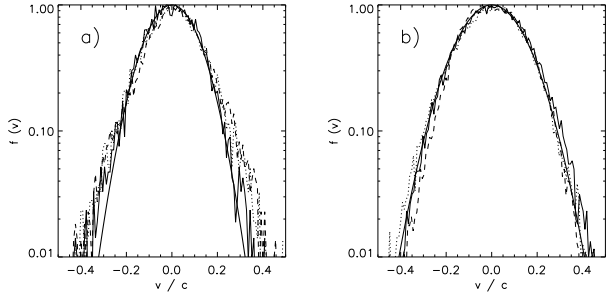}
\end{center}
    \caption{The normalized electron distribution functions
    (thin full line means f(v$_x$), dashed line f(v$_y$), and
    dotted line f(v$_z$)) computed in the regions shown by the white crosses
    in Fig.~\ref{figure3} at two different times: at $\omega_{pe} t$ = 3500
    (a) and at $\omega_{pe} t$ = 7800 (b). The thick full line in (a) means
    the thermal distribution function with the temperature 66~MK
    and the thick full line in (b) the distribution function
    with the temperature 107~MK, which roughly fits the computed
    distribution functions.}
\label{figure4}
\end{figure}

A global evolution of the magnetic field lines and the corresponding electric
current densities (reddish areas) in the system is shown in 
Fig.~\ref{figure2}. When 10~small plasmoids were formed, at about
$\omega_{pe} t = 1200$ (Fig.~\ref{figure2}b), then 
the plasmoids started to interact and merge into larger plasmoids. Due to free
boundaries used in the system and small asymmetries in the initiation, the
plasmoids successively merged into one large plasmoid formed in the bottom part
of the system (Fig.~\ref{figure2}f). On the other hand,
Fig.~\ref{figure3} presents a more detailed 
view on this evolution as well as the distribution of numerical electrons
(points), having the energy greater than 40 keV, at four different times: at
$\omega_{pe} t$ = 3200 (a), at $\omega_{pe} t$ = 3500 (b), at $\omega_{pe} t$ =
5600 (c), and at $\omega_{pe} t$ = 7800 (d). As can be seen here, in each
coalescence process electrons are very efficiently accelerated -- see an
increase of the number of numerical electrons in dependence on time. 
In Fig.~\ref{figure3}b and \ref{figure3}d, we selected the region 
(see white crosses = the centra of the circles
of the radius 70 $\Delta$) with an enhanced number of accelerated electrons. In
these regions we determined the normalized (the maximum equals to 1) electron
distribution functions in all three coordinates
(Fig.~\ref{figure4}a,b). For comparison in 
Fig.~\ref{figure4}a the thermal distribution function  with the
temperature 66~MK is added. 
It shows that this distribution function has clear nonthermal tails. On the
other hand, the computed distribution function in Fig.~\ref{figure4}b
  is nearly the thermal 
one, see the fit of this function by the thick full line corresponding to the
thermal distribution function with the temperature 107~MK.

In summary, Figs.~\ref{figure3} and \ref{figure4} show that the
successive merging (coalescences) of
the plasmoids step by step increase the energy (and the temperature) of
accelerated electrons. We found that at some regions and for short times during
the coalescence process the distribution functions deviate from the thermal
ones (the power-law tails in Fig.~\ref{figure4}a or even
bumps-on-tail, see e.g. Fig.~4 in 
the paper by \citet{KarlickyBarta2007}), but very soon these distribution
functions are thermalized (i.e. changed to the Maxwellian ones) by fast
wave-particle processes (anomalous collisions).

\subsection{Successive merging of plasmoids and the above-loop-top
    hard X-ray sources} 
\label{sect31}

\begin{figure}
\begin{center}
\includegraphics[width=8.5cm]{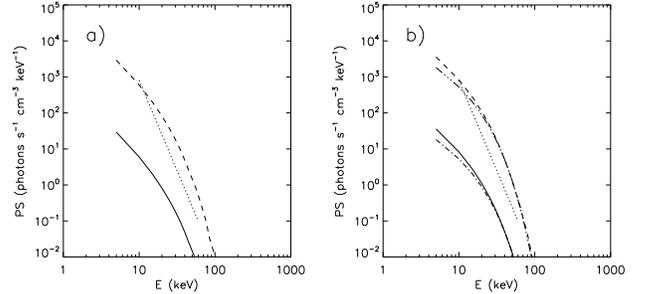}
\end{center}
    \caption{The X-ray spectra (a,b) (the full line for the source density
    n$_e$ = 10$^{9}$ cm$^{-3}$, and the dashed line for n$_e$ =
    10$^{10}$ cm$^{-3}$) 
    corresponding to the distributions in Fig.~\ref{figure4}a, b,
    computed by method (A). 
    The X-ray spectra in (b) (the dashed-dot line for the source density
    n$_e$ = 10$^{9}$ cm$^{-3}$, and the dashed-dot-dot-dot line
    for n$_e$ = 10$^{10}$ cm$^{-3}$)
    corresponding to the distribution in Fig.~\ref{figure4}b, but
    computed by the method (B) for the temperature 107~MK.
    For comparison the X-ray spectrum (dotted line)
    observed during the December 31, 2007 flare (according to Krucker
    et al. 2010) 
    is added.}
\label{figure5}
\end{figure}

We computed the hard X-ray spectra generated at two selected regions (the white
crosses in Fig.~\ref{figure3}) and compared them with that observed
during the December 31, 
2007 flare \citep{Kruckeretal2010}. In accordance with the paper by 
\citet{Kruckeretal2010} we have taken the plasma density 
in the range 10$^{9}$---10$^{10}$
cm$^{-3}$ and the source volume $V$ = 8$\times$ 10$^{26}$ cm$^{3}$. For the
computation of the hard X-ray spectra we used two methods: (A) the non-thermal
bremsstrahlung method, in which the spectrum was computed as a sum of
contributions of the bremsstrahlung emission of all numerical electrons, for
details see the relations (10) and (11) in the paper by 
\citet{KarlickyKosugi2004}, and (B) the thermal 
bremsstrahlung method for specific plasma
temperatures \citep{TandbergEmslie1988}. The computed spectra together
with the observed spectrum are in Fig.~\ref{figure5}a,b. The spectra
computed by the method 
(A) and (B) (in Fig.~\ref{figure5}b) are similar. The observed
spectrum is between the 
spectrum computed for the source plasma density 10$^{9}$ cm$^{-3}$ and that for
10$^{10}$ cm$^{-3}$, especially in later phases of the model evolution. But,
the observed spectrum has a different form, comparing to the computed ones.
Considering our results we think the observed X-ray power-law spectrum is given
by a sum of emissions from many locations with different thermal and nonthermal
distribution functions.

\subsection{Fragmentation between merging plasmoids and narrowband dm-spikes}

\label{sect32}
\begin{figure}
\begin{center}
\includegraphics[width=8.0cm]{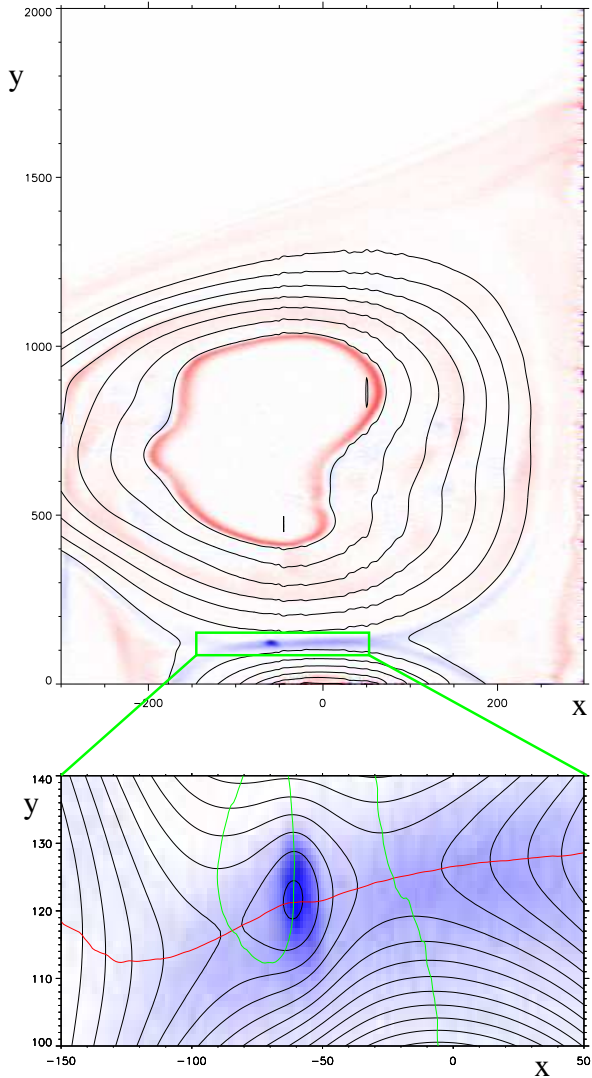}
\end{center}
    \caption{Fragmentation of the current sheet
    in the direction perpendicular to the initial current layer,
    i.e. in the current sheet between two large interacting
    plasmoids, at  $\omega_{pe} t$ = 7200. The reddish areas mean
    the electric current densities with the initial orientation of the
    electric current, 
    and the blue ones mean those with the opposite orientation; red
    and green lines are 
    positions of $B_y = 0$ and $B_x = 0$, respectively; their
    intersections represent 
    the X- and O- type null points.}
\label{figure6}
\end{figure}

In accordance with the results presented by \citet{Bartaal2010b}, we studied
the structure of the magnetic field in the region between merging large
plasmoids. One example of such a structure, formed at the time 
$\omega_{pe} t= 7200$ is presented in Fig.~\ref{figure6}. 
The reddish areas mean the electric current
densities with the initial orientation of the electric current, 
and the blue ones mean those with the opposite orientation; 
red and green lines are
positions of $B_y = 0$ and $B_x = 0$, respectively; their intersections
represent 
the X- and O- type null points. The figure shows that the initial current
sheet is 
fragmented into several sub-current sheets. In these secondary current sheets
further tearing can exist provided they are sufficiently long. Our results thus
indicate, that the fragmentation cascade seen in a large-scale MHD simulation
\citep{Bartaal2010b} continues actually down to the dissipation 
scale of the order $\approx d_i$
(the thickness of the CS at the bottom panel of Fig.~\ref{figure6} is
$\approx 30\Delta$). 
At this scale, however no further fragmentation has been observed, the current
structures dissipate by plasma kinetic processes 'silently'.

\begin{figure}
\begin{center}
\includegraphics[width=6cm]{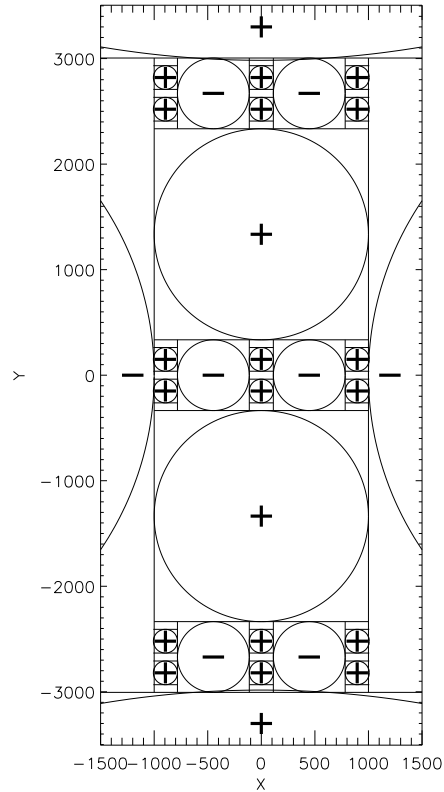}
\end{center}
\caption{Example of the fractal reconnection structure corresponding
to the relations (\ref{eq1}) and (\ref{eq2}), for $A$=2 and $B$=3. Double
straight lines delineate current sheets with interacting plasmoids (circles).
The plus and minus signs express orientation of electric currents in the
plasmoids. $X$ and $Y$ are in arbitrary units.}
\label{figure7}
\end{figure}

Considering the structures of fragmented current sheets (cascade of plasmoids)
obtained by numerical simulations (Fig.~\ref{figure6}), we constructed
the schematic 
structure of fragmentation. An example of such a structure is shown in
Fig.~\ref{figure7}. 
The circles mean plasmoids with positively ($+$) and negatively ($-$) oriented
electric currents, the parallel lines mean boundaries of current sheets. In
such a structure the radius of the plasmoid $R_i$ can be written as:
\begin{equation}\label{eq1}
R_{i} = A R_{i+1} + B R_{i+2},
\end{equation}
where $A$ and $B$ are the numbers of plasmoids and current sheets in specific
current sheet, $i$ means the index of the plasmoids with the same size. The
number of plasmoids increases with $i$ as $n_i \sim$ B$^i$. Then, the ratio
between subsequent radii of plasmoids in the plasmoid cascade can be expressed
by the infinite continued fraction:
\begin{equation}\label{eq2}
\frac{R_{i}}{R_{i+1}} = \left(A + \frac{B}{A + \frac{B}{A + \frac{B}{A +
...}}}\right).
\end{equation}

Now, assuming that the energy in the plasmoids is proportional to their area,
the dependence of the function $E_i$ = R$_i^2$ $\times$ n$_i$ on the k-vector
scale ($k_i$ = 2$\pi$/R$_i$) can be computed. This function is the power-law
one with the power-law index
\begin{equation}\label{eq3}
p = \frac{\log(C\cdot B)}{\log \frac{R_{i}}{R_{i+1}}} - 2,
\end{equation}
where $C$ is the free parameter expressing possible deviations from our
assumption about the plasmoid energy. For the structure presented in
Fig.~\ref{figure7} and 
for $C = 1$ this power-law index is $p = -1.54$.

The power-law dependence were found also in the analysis of the frequency
bandwidth of the narrowband dm-spikes
\citep{Karlicky1996,Karlickyetal2000}.
Therefore, considering this fact, the model scenario shown in 
Fig.~\ref{figure1}, and the
turbulent-plasma model of spikes by \citet{BartaKarlicky2001}, we propose
that the narrowband dm-spikes are generated in fragmentation processes between
merging plasmoids in the reconnection outflow in the region above the flare
loop arcade. While \citet{BartaKarlicky2001} supposed (silently)
MHD/Alfv\'enic turbulence in the super-sonic outflows, our recent simulations
indicate that the tearing/coalescence cascade might be more likely source of
turbulence. However, in a fully developed MHD turbulence all plasma wave-modes
are present, anyway. In the fragmentation region, the plasmoids of all spatial
scales can interact and accelerate electrons. These electrons are trapped in
plasmoids of different sizes. In each plasmoid they can generate the radio
emission in the frequency range corresponding to the range of the plasma
density in this  plasmoid. Due to expected power-law dependence of spatial
scales of these fragmented plasmoids, the dependence of bandwidths of the
resulting radio emission should be the power-law one as observed in the
narrowband dm-spikes.

\begin{figure}
\begin{center}
\includegraphics[width=8.5cm]{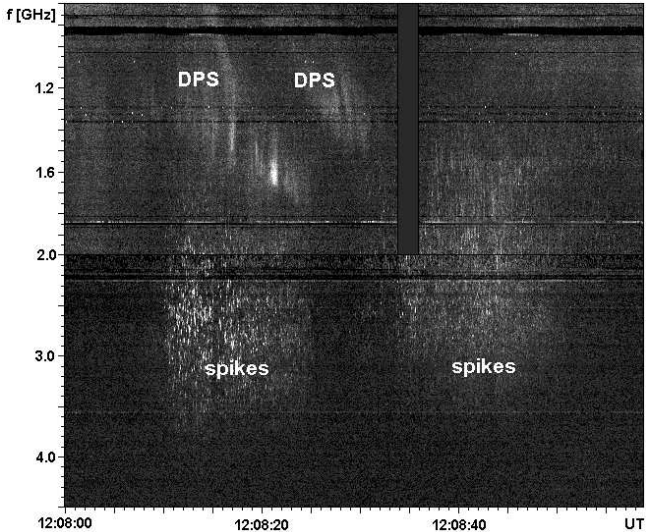}
\end{center}
    \caption{The radio spectrum observed during the 28 March 2001 flare
    by two Ond\v{r}ejov radiospectrographs (0.8--2.0 and 2.0--4.5 GHz)
    supporting our model. It shows the drifting pulsating structures
    (DPSs) which drift towards narrowband dm-spikes. The 0.8--2.0 GHz
    spectrum was shortly interrupted at 12:08:34--12:08:36~UT.}
\label{figure8}
\end{figure}

To support this idea, in Fig.~\ref{figure8} we present the radio
spectrum observed during 
the 28 March 2001 flare by two Ond\v{r}ejov radiospectrographs (0.8---2.0 and
2.0---4.5 GHz) \citep{Jirickaetal1993}. It shows two positively drifting
pulsating structures (DPSs) which according to our previous studies indicate
two plasmoids moving downwards to the sources of the narrowband dm-spikes
(generated in the region of fragmentation, compare to the model scenario in
Fig.~\ref{figure1}). The mean Fourier spectrum of these spikes is the
power-law one with the power-law index $-1.5$.

\section{DISCUSSION AND CONCLUSIONS}

Important aspect of our model is that we used free boundary conditions which
enabled successive merging of small plasmoids into a large final plasmoid. In
simulations, we recognized also a fragmentation of current sheet between two
merging plasmoids.

We showed that these processes very efficiently accelerate electrons to
energies relevant for the emission in the hard X-ray range. Based on this
result we propose that the above-the-loop-top hard X-ray sources are produced
by the successive merging of plasmoids in the region with the turbulent
reconnection outflow, in the region just above the loop arcade. Computed X-ray
spectra support this idea. To explain a difference between the slopes of the
computed and observed X-ray spectra we propose that the observed power-law
spectrum is a sum of emissions from many locations with different thermal and
nonthermal distribution functions.

In simulations we used the particle-in-cell model. Although, the studied
processes are self-similar (i.e. they does not depend on scales, see e.g. the
MHD and PIC simulations in the paper by \citet{Tajimaetal1987}), the results
need 
to be taken with a caution, because the spatial and time scales in the model
and real plasmoids differ in several orders of magnitude. On the other hand, it
is beyond a possibility of any present numerical model to take into account all
these scales.

Further aspect of the PIC modelling is that the Coulomb collisions were not
considered in our model. Namely, the time interval of our computations is much
shorter than the collision time in the-above-the-loop-top X-ray sources; e.g.
$\omega_{pe} t$ = 7800 for the plasma density n$_e$ = 10$^9$ cm$^{-3}$
corresponds to 4.5 $\times$ 10$^{-6}$ s and the collision time to 0.115
T$_7^{3/2}$ s, where T$_7$ = T$_e$/10$^7$. On the other hand, the Coulomb
collisions are essential for the bremsstrahlung X-ray emission. The hard X-ray
spectra in the December 31, 2007 flare were detected as the mean ones over
intervals of several seconds. Furthermore, the-above-the-loop-top hard X-ray
source lasted of about 2 minutes. Thus comparing these times with the collision
one, the collisions influence not only the observed spectra, but also a
duration of the X-ray source. Therefore some re-acceleration of electrons is
even needed. In the present model such a re-acceleration is very probable,
because the plasmoids are generated in a broad range of spatial scales and
electrons can travel several times though regions of interacting plasmoids.
Although an inclusion of the Coulomb collisions and corresponding prolongation
of computations are beyond the possibility of our present PIC modelling, we
think that the observed spectra are given by a competition of the collisions
with the re-acceleration of electrons. On the other hand, in PIC model the
anomalous collisions (wave-particle interactions), which are much more
effective than the Coulomb collisions, are present as confirmed by fast
thermalization of nonthermal distribution functions.

We made additional computations with different initial parameters and we found:
a) the energy gain of accelerated electrons increases with the decrease of the
plasma beta parameter, and b) the increase of the proton-electron mass ratio
m$_p$/m$_e$ makes computations longer, but results are similar.

We compared the present simulation also with that in the numerical model which
size was two times smaller ($L_x \times L_y$ = 600$\Delta$ $\times$
2000$\Delta$) and in which only 5 plasmoids  were initiated (contrary to 10
plasmoids in the present simulation). In this case the final mean energy of
accelerated electrons was 5.3 times greater than the initial one, compare with
that of 10.7 times (from the initial temperature 10~MK to final temperature 
107~MK) in the present case. Namely, each coalescence process increases the
energy 
of accelerated electrons, therefore the number of successive coalescence
processes is essential for their final energy.

For calculations of the hard X-ray spectra, presented in
Fig.~\ref{figure5}b we used two
methods. The obtained results are similar. Small differences  are  due to
differences in these methods and deviations of the computed distribution
functions from the thermal one.

The plasmoids in 2-D are in reality 3-D magnetic ropes. While in 2-D the
trapping of energetic electrons is a natural consequence of a close magnetic
field structure of the plasmoid, in 3-D, this structure is only semi-closed.
However, we consider the merging processes in the turbulent reconnection
outflow therefore the magnetic trapping of electrons, similar to that proposed
by \citet{Jakimiec1998} is highly probable. Moreover, the coalescence
fragmentation process, which generates the reverse electric currents (which in
3-D has to be closed in finite volume) will contribute to a full trapping of
electrons.

In agreement with the conclusions by \citet{Kruckeretal2010}, in the model the
acceleration region is very close to the hard X-ray source. It enables to
re-accelerate energetic electrons, which loss their energy due to collisions.
Acceleration regions are among interacting plasmoids and also between the
plasmoids and the arcade of flaring loops. This model can explain not only the
above-the-loop-top hard X-ray sources, but also the loop-top sources because
the arcade of loops is, in principle, the 'plasmoid' fixed in its half height
at the photosphere.

Considering all aspects of the fragmentation process (power-law spatial scales
of plasmoids, effective acceleration of electrons, trapping of electrons in
plasmoids, location in the reconnection plasma outflow) we think that this
process can explain a generation of the narrowband dm-spikes. We supported this
idea by the radio spectrum observed during the 28 March 2001 showing DPSs
drifting towards the narrowband dm-spikes. Furthermore, it is known that more
than 70~$\%$ of all groups of dm-spikes are observed during the
GOES-rising-flare phases \citep{Jirickaeta2001}. Although these
arguments support the presented idea, further analysis of the narrowband
dm-spikes and their modelling is necessary.

In this first study, we considered only the neutral current sheet, 
i.e. $B_z=0$.
For more realistic description, we plan to extend our study also to cases
with non-zero guiding magnetic field.

The presented model is a natural extension of our previous models explaining
the plasmoid formation, its ejection, and corresponding DPS. The question
arises why the above-the-loop-top hard X-ray sources are very rare comparing,
e.g., with DPSs or dm-spikes observations. We think that the above-the-loop-top
hard X-ray source is large and stationary plasmoid, which is sufficiently dense
and in which there is sufficient amount of energetic electrons. Positional
stationarity and location of this plasmoid is given by surrounding magnetic
field and location where this plasmoid starts to be formed, see the paper by
\citet{Bartaal2008b}. On the other hand, the plasmoids generating DPSs or
dm-spikes need not to be so dense and they need not to have such
amount of energetic
electrons. It is known that for generation of the radio emission a number of
energetic electrons can be much smaller than that for the hard X-ray emission.
Namely, the intensity of the radio emission depends on derivatives of the
electron distribution function in the momentum space; not on the absolute
amount of energetic electrons as in the case of the hard X-ray emission.

\acknowledgements This research was supported by Grant IAA300030701 of the
Grant Agency of the Academy of Sciences of the Czech Republic and the
research project AV0Z10030501 of Astronomical Institute of the Czech
Academy of Science.

\end{document}